\documentstyle [12pt]{article}
\title{
\hspace{3.0truein}{\small IFT-494-UNC}\\
\vspace{0.2truein}
{Topological Invariance at Finite Temperature}
}
\author{Wei Chen\footnotemark[1]\\
Department of Physics\\
University of North Carolina\\
Chapel Hill, NC 27599-3255}
\date{}
\footnotetext{$^*$ chen@physics.unc.edu}

\begin{document}
\maketitle
\vspace{0.2truein}
\begin{abstract}
We examine the thermal behavior of a
theory with charged massive vector matter coupled to
Chern-Simons gauge field. We obtain a critical
temperature $T_c$, at which the effective mass of vector
field vanishes, and the system transfers from a
symmetry broken phase to topological phase.
The phase transition is suggested to be of the zeroth order,
as the free energy of the system is discontinuous at $T_c$.
Application to the $(2+1)$ dimensional
quantum gravity is briefly discussed.
\end{abstract}

\newpage

\renewcommand{\theequation}{\thesection.\arabic{equation}}
\baselineskip=18.0truept
\vskip 0.5truein
\section{Introduction}
\setcounter{equation}{0}
\vspace{3 pt}

Symmetries and symmetry breaking
serve as a basic principle
in modern physics. For instance, it is generally
appreciated now that all the four basic interactions in
Nature can be formulated as theories with
gauge invariance.
However, there are still more to be understood,
such as how
gravity may be quantized, as quantum gravity is believed, for instance,
to have dominated for a period (thought  short)
after big bang and determined the evolution  of
Universe. There seems a need for larger symmetries
to set up a framework for quantum gravity.
There has been argument that a fundamental theory of quantum
gravity will not involve a space-time metric \cite{Pe}.
A theory depending not on a choice of
space-time metric is invariant under
diffeomorphism transformation, and is known as topological
theory.
This topological symmetry rules out any local dynamical degree of
freedom, and the number of states of such a system
could be as few as one - the vacuum - only.

Topological quantum field theories are
studied in the context of low dimensional topology
\cite{A}. In particular, the topological
quantum Yang-Mill theory is used to interpret
the Donaldson theory
which is regarded as a key to understanding the four
dimensional topology, and thus as a key to
understanding the geometry
structure of the space-time \cite{W1}.
In the three dimensions, the topological quantum
Chern-Simons gauge theory is shown
to provide a natural framework of
studying the Jones polynomial of knot theory \cite{W2},
and to have important implications
for two dimensional rational
conformal field theories \cite{W2}\cite{MS}.

Progress is also made in
understanding quantum gravity.
It is shown that $(2+1)$ dimensional
general relativity
is equivalent to a topological field theory
- the Chern-Simons gauge theory
for the Poincare group, and thus, as a quantum
field theory, is exactly soluble   \cite{W3}.
Various topological models relating to gravity in
different spacetime dimensions have been suggested in the
literature \cite{Ts}\cite{DJT}\cite{Ho}.
On the other hand, since any theory one uses
to describe physics
at low energy and/or low temperature
involves a spacetime metric and local dynamical
degrees of freedom,
there must be some means that connects a
topological theory and its low
energy/temperature limit. In particular,
if a topological quantum field theory describes
a phase of unbroken diffeomorphism of
quantum gravity, a metric and local dynamics might
arise as some form of symmetry breaking.
Such a possible topological phase transition might
happen in the early stage of Universe
when it was extremely hot and small in size.

In this Paper, following a recent work \cite{CH}, we
examine a system that exhibits a topological phase transition.
The system we work with is a $(2+1)$ dimensional one
with a charged massive vector matter coupled to a $U(1)$ gauge
field. Both the vector and gauge fields are governed by
a Chern-Simons term. While  the Chern-Simons
gauge field does not carry local dynamical degree of freedom (and
is known as a topological field),
the vector matter field carries them 
due to its mass term. Moreover, as the mass term involves
a space-time metric, the massive vector field theory and thus
the interacting theory are obviously topologically variant.
Now, however, if somehow the vector field
becomes massless, the local dynamical degrees of freedom, the space-time
metric, and the difference between the gauge and matter fields disappear.
And  then the system , now with non-Abelian
gauge symmetry, turns out to be topologically invariant.
Namely, a topological phase transition would happen. The mass of
vector field naturally serves as
the order parameter characterizing the possible transition.
To realize it, we heat up
the system and see there indeed exists a critical temperature at
which the effective vector mass vanishes.

Breaking and restoration of various
symmetries at high energy and/or finite
temperature have been known
for a long time in quantum field theories.
Notable theoretical observations include
the finite temperature restorations of
$SU(2)\times U(1)$ gauge symmetry for the
weak electric interactions 
\cite{L}\cite{DJ}\cite{W} and
the deconfinement transition in the quark-gluon
plasma (for a review, see for instance \cite{GPY}).
Here we provide an example of phase transition
concerning topological and gauge symmetries.
The transition, recognized as of the zeroth order,
is thus far as we know the most discontinuous -
even the free energy density
discontinues  at the critical temperature.
Our conclusions are drawn based on perturbation expansion
over the Chern-Simons coupling to the second order.
We also observe that, with the thermal fluctuation,
the effective Chern-Simons coupling
in the symmetry broken phase is
just a slowly increasing function of temperature $T$ with $g(T=0) = g_r$ and
$g(T=T_c) = 2g_r$ at the second order, as we shall see below. Therefore
the perturbation expansion and our conclusions should be reliable
in the whole range of temperature $T$, as long as the renormalized
zero temperature Chern-Simons coupling $g_r$ is small.

In the next section,
we review the model in some details with emphasis on symmetries.
Section 3 is for deriving the partition function
in path integral formalism. In section 4, we calculate one
loop corrections to two- and three-point
correlation functions. From these we obtain the critical temperature
for the transition, and verify the
reliability of perturbation expansion.
We then calculate in section 5
free energy density to two loops.
{}From it, the order of topological
phase transition is determined. Section 6 is devoted to
summarizing our results.

\renewcommand{\theequation}{\thesection.\arabic{equation}}
\vskip 0.5truein
\section{The Model}
\setcounter{equation}{0}
\vspace{3 pt}

The model of interests is a massive complex vector field $B_\mu$
minimally coupled to a gauge field $a_\mu$ who's dynamics is
governed by a Chern-Simons term,
\begin{equation}
I = \int_\Omega\frac{1}{2}\left(
\epsilon_{\mu\nu\lambda}
B^*_\mu(\partial_\nu - iga_\nu)B_\lambda
+MB^*_\mu B^\mu
+ \epsilon_{\mu\nu\lambda}
a_\mu\partial_\nu a_\lambda\right)\;,
\label{S}
\end{equation}
where the three space-time manifold $\Omega$ has Lorentzian signature.
This theory was studied in a different context (anyon)
\cite{CI1}\cite{JN}.
This is an Abelian gauge field theory, violating parity and
time reversal symmetries. Though the $B_\mu$ field involves
a Chern-Simons term as well, its nature is rather different
from that of $a_\mu$ field,
because of its non-vanishing mass term.
Under a $U(1)$ transformation
parametrized by $\alpha({\bf x},t)$, the Chern-Simons field
$a_\mu$  varies as a gauge field
\begin{equation}
a_\mu ({\bf x},t) \rightarrow a_\mu ({\bf x},t)
+ \partial_\mu \alpha ({\bf x},t)\;,\label{t1}
\end{equation}
and the massive $B_\mu$ field transforms like a charged matter field
\begin{equation}
B_\mu ({\bf x},t) \rightarrow e^{i\alpha({\bf x},t)}B_\mu({\bf x},t)\;.
\label{t2}
\end{equation}
As a consequence of the $U(1)$ symmetry,
the Chern-Simons gauge field $a_\mu$ does not carry dynamical
degree of freedom, but the massive vector field $B_\mu$ does.
To see this, we consider the equations of motion. For the Chern-Simons
field $a_\mu$, it is
\begin{equation}
\epsilon_{\mu\nu\lambda}\partial_\nu a_\lambda =
j_\mu\;,
\label{eq1}
\end{equation}
where the current $j_\mu
= -i\epsilon_{\mu\nu\lambda}B^*_\nu B_\lambda$.
{}From (\ref{eq1}), one obtains the current conservation
\begin{equation}
\partial_\mu j_\mu = 0\;.
\end{equation}
Moreover, from (\ref{eq1}), one can solve $a_\mu$
by expressing it in integrals over the current
$j_\mu$, with a use of gauge fixing to $a_\mu$.
This implies the Chern-Simons field is
completely determined by the matter
field. On the other hand, the equation of motion for
the massive vector field $B_\mu$ is
\begin{equation}
\epsilon_{\mu\nu\lambda}(\partial_\nu - iga_\nu)
B_\lambda +iMB_\mu = 0\;.\label{eq2}
\end{equation}
Acting $(\partial_\mu-iga_\mu)$ on (\ref{eq2}), for $M\neq 0$,
one obtains
\begin{equation}
(\partial_\mu - iga_\mu)B_\mu = 0\;.
\label{con1}
\end{equation}
As $a_\mu$ is not an independent field, (\ref{con1}) ensures that
one of the three components of $B_\mu$ is eliminated in a covariant way.
Another way to see the difference between $a_\mu$ and
$B_\mu$ fields
is via the canonical approach, in which an independent
degree of freedom must be accompanied by
a non-vanishing canonical momentum
conjugate, and vise versa.  Let's
write down the canonical momentum conjugate of the
field variable $a_\mu$
\begin{equation}
\pi_\mu=\frac{\delta{\cal L}}{\delta \dot{a}_\mu}
= \frac{1}{2}\epsilon_{0\mu\lambda}a_\lambda\;.
\label{cj}
\end{equation}
Now one sees $a_0$ is not an independent degree of
freedom as $\pi_0 = 0$. Moreover, the gauge transformation
(\ref{t1}) implies one gauge degree of freedom of $a_\mu$
should be fixed. A simple choice is the axil gauge $a_2=0$.
Then, as $a_1$'s conjugate
$\pi_1 = \frac{1}{2}a_2 = 0$ [see (\ref{cj})], $a_1$ is not an independent
degree of freedom either.
Therefore, the Chern-Simons field $a_\mu$ has no local
dynamical degree of freedom at all.
On the other hand, the canonical momentum conjugate of
the massive $B_\mu$ field takes the same form as given in
(\ref{cj}).  Therefore $B_0$ is not an independent field variable.
However, there is no gauge freedom
associated to the massive $B_\mu$ field, and
$B_1$ is  the independent field variable and $B_2$ the canonical
momentum conjugate, or vise versa.

On the other hand, when $M=0$, the difference
between $a_\mu$ and $B_\mu$ disappears, except the
latter is complex. Then,
for convenience,
we set $ gB_\mu = A_\mu^1+iA^2_\mu$ and
$ga_\mu=A^0_\mu$.
Substituting these into (\ref{S}),
we arrive at an action
\begin{equation}
I_{CS} = \frac{1}{8\pi\alpha}\int_\Omega
\epsilon_{\mu\nu\lambda}
\left(A^a_\mu\partial_\nu A^a_\lambda
+ \frac{1}{3}\epsilon^{abc}A^a_\mu A^b_\nu
A^c_\lambda\right)\;,
\label{Scs}
\end{equation}
where $g^2=4\pi\alpha$. Carrying no local dynamical degree of
freedom,  this is known as a topological
field theory.
Now, the gauge invariance of the theory is not simply $U(1)$,
but a non-Abelian one. The gauge group can be $SU(2)$ or $ISO(2,1)$,
with structure constant $\epsilon^{abc}$.
The variation of $A^a_\mu$ under a gauge
transformation is
\begin{equation}
\delta A_\mu^a = D_\mu \tau^a\;,
\label{gt1}
\end{equation}
where $D_\mu\tau^a = \partial_\mu\tau^a
+\epsilon^{abc}A^b_\mu\tau^c$.

Moreover, independent of a choice of
spacetime metric, (\ref{Scs}) is
invariant under diffeomorphism
transformations. Such a transformation can be generated by a
vector on the three manifold $\Omega$,
 $V^\mu$, via Lie derivative ${\cal L}_V$.
Under it, the Chern-Simons field $A_\mu$ transforms as
 \begin{equation}
{\cal L}_VA_\mu^a = D_\mu(A^a_\nu V^\nu) + V^\nu F^a_{\nu\mu}\;,
\label{dif}
\end{equation}
so that
\begin{equation}
{\cal L}_VI_{CS}[A] = 0\;,
\end{equation}
provided $V^\mu$ vanishes on the boundaries of  $\Omega$
or the space-time manifold has no boundary
\cite{Ho}\cite{Chen}.
However, subject to the flat connection condition
$F_{\mu\nu}= [ D_\mu, ~D_\nu ] = 0$,
diffeomorphism transformations
completely fall into gauge transformations. In fact,
the first term of
(\ref{dif}) can be identified as the gauge transformation
(\ref{gt1}) with gauge parameter $\tau^a = A^a_\nu V^\nu$,
and the second term is proportional to the
flat connection condition. Therefore, on the constraint surface,
a generator generates simultaneously both
diffeomorphism and gauge transformation. Consequently,
a gauge choice fixes both the gauge and diffeomorphism.
As a topological quantum field theory,
(\ref{Scs}) is exactly soluble \cite{W2}.

Now, we see that the mass parameter $M$ plays the role
of order parameter. When $M > 0$, the system described
by (\ref{S}) has  $U(1)$ symmetry only; however when $M=0$,
the system is invariant under non-Abelian gauge and topological
transformations.
In the following sections, we shall consider the
quantization of the massive theory, {i.e.} the system in a symmetry
broken phase, in a heat bath.
We shall see then that the mass of the vector field $B_\mu$
becomes temperature
dependent, $M=M(T)$, and there exists a critical temperature $T_c$ at
which $M(T) = 0$, and a topological phase transition
of the zeroth order happens.

Another parameter of the theory is the Chern-Simons coupling $g$.
It is dimensionless as the Chern-Simons interaction operator is
marginal. This implies that the Chern-Simons coupling may need
potentially non-trivial renormalization.
However, due to the topological nature of the Chern-Simons term,
$g$ is not sensitive to changes of energy scale. In other words,
the beta function of $g$ vanishes identically, and $g$ is not
a running coupling constant. This fact makes a 'small'
Chern-Simons coupling a perfect controlling
parameter in perturbation expansion. On the other hand, however,
when the thermal effect is taken into account, the concern is
now that the Chern-Simons coupling may be temperature dependent,
and if it went up rapidly with temperature,
perturbation would break down. We shall see below that fortunately
it is not the case. The effective (temperature dependent) Chern-Simons
coupling
is just a slowly increasing function of temperature, and at the
critical temperature it is only
twice as large as the renormalized zero temperature one.

\renewcommand{\theequation}{\thesection.\arabic{equation}}
\vskip 0.5truein
\section{Partition Function}
\setcounter{equation}{0}
\vspace{3 pt}

Now we attach the system to a heat bath with temperature $T$.
As is known, finite temperature behavior of any theory
is specified by the partition function
\begin{equation}
Z={\rm Tr}e^{-\beta H}\;;
\end{equation}
and the thermal expectations of physical observables
\begin{equation}
<{\cal O}> ~= \frac{1}{Z}
{\rm Tr}[{\cal O}e^{-\beta H}]\;,
\end{equation}
where $\beta = 1/T$ is the inverse temperature ($k_B = 1$) \cite{K}.

As our system is relativistic, we like to
work out its covariant formalism.
It is clear from the discussion in the previous section,
the Chern-Simons theory involves
both the primary (first class)
and secondary  (second class)
constraints. Therefore, to derive the partition function in
the path integral formalism, special cares are needed.
We shall outline the derivation in the zero temperature
field theory \cite{AB}, and then switch to the finite
temperature case. We shall focus on
the Chern-Simons field $a_\mu$, {\it i.e.} consider the theory
\begin{equation}
{\cal L}_{cs}=
-\frac{1}{2}\epsilon_{\mu\nu\lambda}
a_\mu\partial_\nu a_\lambda
+a_\mu j_\mu\;.
\label{cs}
\end{equation}
Generalization to
the massive field $B_\mu$ will be then straightforward.
With the momentum conjugates $\pi_\mu$ given in (\ref{cj}),
it is easy to check that the theory (\ref{cs})
is subject to the following
constraints:
\begin{eqnarray}
\chi_0 &=&\pi_0\;,\\
\chi_i &=&\pi_i - \frac{1}{2}\epsilon_{ij}a_j \;,\\
\chi &=&\epsilon_{ij}\partial_ia_j+j_0\;.
\end{eqnarray}
The last one is from
$\delta{\cal L}_{cs}/\delta a_0 = 0$. Then
the total Hamiltonian density is
\begin{equation}
{\cal H} = {\cal H}_c
+ \lambda_0 \chi_0 + \lambda_i \chi_i
+ \lambda \chi \;,
\label{h1}
\end{equation}
where $\lambda$'s are the lagrange multipliers, and the canonical Hamiltonian
density  ${\cal H}_c$ is
\begin{eqnarray}
 {\cal H}_c &=& \dot{a}_\mu\pi_\mu - {\cal L}_{cs}\;\\
&=& \dot{a}_0\pi_0+ \dot{a}_i(\pi_i-\frac{1}{2}\epsilon_{ij}a_j)
+ {a}_0(\epsilon_{ij}\partial_ia_j+j_0)
- a_ij_i\;.
\label{hc}
\end{eqnarray}
{}From (\ref{h1}) and (\ref{hc}),
one sees that $\dot{a}_0$,  $\dot{a}_i$ and $a_0$ in ${\cal H}$ can be absorbed
by the lagrange multiplier $\lambda_0$, $\lambda_i$ and $\lambda$,
respectively. Doing so, $a_0$ disappears, and so one can
discard $\pi_0$ by setting $\lambda_0 = 0$. The Hamiltonian density is
now
\begin{equation}
{\cal H} = \lambda_i \chi_i + \lambda\chi -a_ij_i\;.
\end{equation}
Now two constraints remain. The first,
 $\chi_i$, is of second class.
And the second, $\chi$,
seems  to be of second class too. However,
upon a linear combination with
$\chi_i$, a redefined constraint,
$\partial_i\pi_i +\epsilon_{ij}\partial_ia_j+j_0$,
turns out to be of first class. As a matter of fact,
this first class constraint is due to the gauge freedom of
the Chern-Simons field.
 A nice approach to deal with a quantum system
with first and second class constraints is to
replace the Poison brackets with the
Dirac brackets \cite{Di} so that the second class constraints can
 be absorbed into the measure of the path integrals \cite{Se}
and the first class ones are taken care by using gauge fixing terms.
For this purpose, let us choose a fixed time $t$. The constraint matrix
element $C_{ij}$ at the time $t$ is
\begin{equation}
C_{ij}({\bf x},{\bf y}) = \{\chi_i({\bf x}),
\chi_j({\bf y})\}
= - 2\epsilon_{ij}\delta({\bf x}-{\bf y})\;;
\end{equation}
and the inverse
\begin{equation}
C_{ij}^{-1}({\bf x},{\bf y})
= \frac{1}{2}\epsilon_{ij}\delta({\bf x}-{\bf y})\;.
\end{equation}
Then the basic (equal time) Dirac brackets
(also called star brackets) are
\begin{eqnarray}
\{a_i({\bf x}),a_j({\bf y})\}_D
&=&  \frac{1}{2}\epsilon_{ij}\delta({\bf x}-{\bf y})\;,\\
\{\pi_i({\bf x}),\pi_j({\bf y})\}_D
&=&  2\epsilon_{ij}\delta({\bf x}-{\bf y})\;,\\
\{a_i({\bf x}),\pi_j({\bf y})\}_D
&=& \eta_{ij}\delta({\bf x}-{\bf y})\;.
\end{eqnarray}
And the first class constaint satisfies
\begin{equation}
 \{\chi({\bf x}),\chi({\bf y})\}_D = 0\;.
\end{equation}
Now,   to fix the first class constraint $\chi$,
one introduces a pair of ghosts,
$c$ and $\bar{c}$,
and their momentum conjugates, $\bar{f}$ and $f$,
and the momentum conjugate of the lagrange multiplier $\lambda$,
$\sigma$. With these new fields, the
system has the
global nilpotent (off-shell) $BRST$
symmetry
\cite{BRST}\cite{Chen}. The $BRST$ charge
\begin{equation}
Q = \int d^2x (c\chi - i f\sigma)\;
\end{equation}
generates the $BRST$ transformations for all the fields
via the Dirac brackets.
Now, the partition function is
\begin{equation}
Z_{cs} = N\int[{\cal D}\mu] e^{iI_{eff}}\;,
\end{equation}
with
\begin{equation}
I_{eff} = \int d^3x
\left(\dot{a}_i\pi_i
+\dot{\lambda}\sigma + \dot{c}\bar{f}
+\dot{\bar{c}}f +a_ij_i
+ \{\Phi, Q\}_D\right)\;,
\end{equation}
where $\Phi$ is a gauge fixing function, and the measure
\begin{equation}
[{\cal D}\mu] =
({\rm det}|{\chi_i,\chi_j}|)^{1/2}
\prod_i\delta(\chi_i)
{\cal D}a_i{\cal D}\pi_i{\cal D}
\lambda{\cal D}\sigma{\cal D}c{\cal D}
\bar{c}{\cal D}f{\cal D}\bar{f}\;.
\end{equation}
The determinant factor in the measure can be absorbed into the
normalization factor $N$. And the integrations over
$\pi_i$ eliminate the second class constraint
$\chi_i$. Moreover, as a covariant
gauge fixing for the later use,
the gauge function is so chosen
\begin{equation}
\Phi = i\bar{c}\left(\frac{1}{2}\rho \sigma -\partial_i a_i\right)
+ \bar{f}\lambda\;
\end{equation}
 that
\begin{equation}
\{\Phi, Q\}_D = -\frac{\rho}{2}\sigma^2-i\bar{c}\partial_i\partial_ic +
\sigma \partial_ia_i -i \bar{f}f
-\lambda(j_0+2\epsilon_{ij}\partial_i a_j)\;,
\end{equation}
where $\rho$ with dimension in the unit of mass parametrizes
the covariant gauge fixing for $U(1)$ symmetry.
Now, integrating out all the momenta
($\sigma$, $\bar{f}$,
$f$ and $\pi_i$) and denoting the lagrange multiplier
$\lambda$ as $a_0$ in the partition
function, we obtain finally
\begin{equation}
Z_{cs}=N\int{\cal D}a_\mu{\cal D}c{\cal D}\bar{c}
{\rm exp}\left(i\int d^3x(\frac{1}{2}\epsilon_{\mu\nu\lambda}
a_\mu\partial_\nu a_\lambda + a_\mu j_\mu +
i\bar{c}\partial^2c
+ \frac{1}{2\rho}(\partial_\mu a_\mu)^2)\right)\;.
\end{equation}
Above, the
ghost fields $c$ and $\bar{c}$
do not interact
with other fields but only serve to
cancel the non-physical, in fact all,
degrees of freedom of The Chern-Simons field $a_\mu$.

The derivation of the partition function for
the vector $B_\mu$ field can be
 done similarly. The only difference is
that there is no gauge freedom and so
no first class constraint for
the $B_\mu$ field, since it is massive.

With these preparation, it is readily to work out the
functional integral representation of
partition function at finite temperature $T$.
The trick is rather simple: to
replace the time variable $t$ with
the imaginary time $i\tau$ via a Wick rotation,
and to explain the final imaginary time
as the inverse temperature $\beta = 1/T$.
Then the partition function of system described by
(\ref{S}) is
\begin{equation}
Z = N\int\prod_\mu\prod_\nu\prod_\lambda{\cal D}a_\mu
{\cal D}B^*_\nu
{\cal D}B_\lambda
{\cal D}c
{\cal D}\bar{c}
{\rm exp}\left( -\int_0^\beta d\tau\int d^2x
{\cal L}\right)\;,
\label{Z}
\end{equation}
with the Euclidean Lagrangian
\begin{equation}
{\cal L}=
-\frac{i}{2}\epsilon_{\mu\nu\lambda}
B^*_\mu(\partial_\nu - iga_\nu)B_\lambda
+\frac{M}{2}B^*_\mu B_\mu
-\frac{i}{2}\epsilon_{\mu\nu\lambda}
a_\mu\partial_\nu a_\lambda
+ (\partial_\mu\bar{c})(\partial_\mu c)
+\frac{1}{2\rho}(\partial_\mu a_\mu)^2
\;.
\label{Se}
\end{equation}
Being vector bosons or ghosts with ghost number
$\pm 1$, all fields in (\ref{Z}) are subject to the
periodic boundary condition such as
\begin{equation}
a_\mu(\beta, {\bf x})=a_\mu(0, {\bf x})
{}~~~~ {\rm and} ~~~~B_\mu(\beta, {\bf x})=B_\mu(0, {\bf x})\;.
\label{bc}
\end{equation}

{}From the partition function (\ref{Z}),
it is easy to work out the finite temperature
Feynman rules. The Chern-Simons propagator in the
Landau gauge and the vertex are
\begin{equation}
D^0_{\mu\nu}(p)
= \frac{\epsilon_{\mu\nu\lambda}p_\lambda}{p^2}\;,
{}~~~~{\rm and} ~~~~
\Gamma^0_{\mu\nu\lambda}=g\epsilon_{\mu\nu\lambda}.
\label{R}
\end{equation}
And the vector propagator is
\begin{equation}
{}~G^0_{\mu\nu} = \frac{\epsilon_{\mu\nu\lambda}p_\lambda
+\delta_{\mu\nu}M+p_\mu p_\nu/M}{p^2+M^2}\;.
\label{R1}
\end{equation}

According to the periodic boundary condition (\ref{bc}),
the third component of
momentum, the frequency, takes discrete values,
$p_3 = 2\pi nT$ for integer $n$'s.
Besides, each loop in a Feynman diagram
carries an integration-summation $T\sum_n\int d^2p/(2\pi)^2$
over the
internal momentum-frequency $({\bf p},2\pi Tn)$; and at each
vertex, the momentum-frequency conservation is satisfied.

\vskip 0.5truein
\section{Critical Temperature}
\setcounter{equation}{0}
\vspace{3 pt}

We now consider perturbation expansion of the theory. In this
section, the two and three point correlation functions are calculated
to the second order. For our purpose, we shall set the
external momenta-frequencies zero.
To start, we cite a useful equation that maps the
discrete summation
$T\sum^{\infty}_{n=-\infty}f(p_3=2\pi Tn)$ into continuous integrals.
By a contour integral on a complex plane, one obtains
\begin{equation}
T\sum^{\infty}_{n=-\infty}f(p_3=2\pi Tn)=
\frac{1}{2}\int_{-\infty}^\infty \frac{dz}{2\pi}[f(z)+f(-z)]
+\int_{-\infty+i\epsilon}^{\infty+i\epsilon}
\frac{dz}{2\pi}[f(z)+f(-z)]\frac{1}{e^{-i\beta z}-1}\;.
\label{f}
\end{equation}
The expression is correct as long as the function $f(p_3)$ has
no singularities along the real $p_3$ axis. One advantage of
this equation is that it naturally separates the
the temperature independent piece from the temperature dependent
one. As an application,
let`s consider a typical integration-summation
\begin{equation}
J(M,T)=T\sum^\infty_{n=-\infty}\int\frac{d^2p}{(2\pi)^2}\frac{1}{p^2+M^2}\;,
\end{equation}
which will appear in loop calculations.
By using (\ref{f}), the frequency sum is converted into the integrals
\begin{equation}
J(M,T)=\int\frac{d^3p}{(2\pi)^3}\frac{1}{p^2+M^2} +
2\int_{-\infty+i\epsilon}^{\infty+i\epsilon}
\frac{dz}{2\pi}\int\frac{d^2p}{(2\pi)^2}
\frac{1}{z^2+{\bf p}^2+M^2}\frac{1}{e^{-i\beta z}-1}\;.
\label{J}
\end{equation}
The first term above is ultraviolet divergent and a regularization is
necessary. If a naive cutoff $\Lambda$ is introduced, the result is
\begin{equation}
\int_\Lambda\frac{d^3p}{(2\pi)^3}\frac{1}{p^2+M^2}
=\frac{\Lambda}{2\pi^2} - \frac{M}{4\pi}\;.
\label{ze}
\end{equation}
On the other hand, if a gauge invariance reserved regularization such
as the dimensional regularization is used, one
ends up with the second term in (\ref{ze}) only, though the integral
is power-counting linear divergent in three dimensions.
Namely, the two regularization
schemes differ one another only by a $\Lambda$-dependent term.
This sort of linearly cutoff dependent terms can be absorbed by
re-definition of the zero temperature mass and coupling constant,
as we shall see below, and therefore no physics should be affected
by the regularization procedure(s) used.
The second term in (\ref{J}) involves no divergent, thanks to the
Bose-Einstein distribution function. This is a well-known
feature of the finite temperature theory 
that all the divergences appear and thus can be taken care in the zero
temperature field theories, and thermal effects are finite
at any finite temperature $T$ \cite{Wein}\cite{MK}.
The integrals on the complex
$z$ plane and on the real two-dimensional ${\bf p}$ space in the second term
of (\ref{J}) are readily
to perform, and we obtain
\begin{equation}
J(M,T) = \frac{\Lambda}{2\pi^2} - \frac{M}{4\pi}
-\frac{T}{2\pi}ln(1-e^{-M/T})\;.
\label{J1}
\end{equation}

Now we consider one loop corrections
to the correlation functions with vanishing
external momenta.
The inverse two-point function of $B_\mu$
field is defined as
\begin{equation}
{G_{\mu\nu}}^{-1}(p) = {G^0_{\mu\nu}}^{-1}(p)
-\Sigma_{\mu\nu}(p)\;,
\end{equation}
where $\Sigma_{\mu\nu}(p)$ denoting  the self-energy. Then
the effective mass of the vector field $B_\mu$
can be defined as 
\begin{equation}
M(g,T)\delta_{\mu\nu}
= {G_{\mu\nu}}^{-1}(p=0)\;.
\label{Mt}
\end{equation}
In general,
with only $O(2)$ symmetry in the space,
the effective mass $M(g,T)$ of a vector field in the longitudinal
direction is not necessarily the same with that in the transverse
direction. For the massive vector field we are considering,
it is the case.

It is not difficult to check that the tadpole diagrams have not
contributions to the self-energy. Then calculating
\begin{equation}
\Sigma^{(2)}_{\mu\nu}(p) = g^2T\sum_n\int\frac{d^2q}{(2\pi)^2}
\epsilon_{\mu\sigma\eta}
G^0_{\sigma\lambda}(p+q)\epsilon_{\lambda\tau\nu}D^0_{\tau\eta}(q)\;,
\label{Sigma}
\end{equation}
we have
\begin{equation}
\Sigma^{(2)}_{\mu\nu}(p=0) = \frac{2}{3}g^2\delta_{\mu\nu}J(M,T)\;,
\end{equation}
and the effective mass by using (\ref{Mt}) 
\begin{equation}
M(g_r,T) =M_r+
\frac{1}{6\pi}g_r^2 
\left(M_r
+ 2T{\rm ln}(1 - e^{-M_r/T})\right) + {\cal O}
(g^4)\;,
\label{MT}
\end{equation}
where $M_r$ denotes  the renormalized zero temperature
mass. For instance, $M_r = M-\frac{1}{3\pi^2}g^2\Lambda$
at one-loop in the regularization by a naive ultraviolet cutoff
$\Lambda$, or $M_r = M$ in the dimensional regularization.
In the bracket in (\ref{MT}), the bare mass $M$ has been replaced by
the renormalized $M_r$, and the bare zero temperature
Chern-Simons coupling $g$ by the renormalized $g_r$ (see below),
 this replacement affects at most higher orders.
The first term in the bracket of (\ref{MT})
is the radiative correction at zero temperature.
Without a symmetry to restrict the renormalized vector mass to
particular values, especially zero, the vector mass is a free
parameter. Indeed, as we have seen here that, even one starts
with a massless theory $M=0$, the radiation
correction at zero temperature generates a vector mass via
the Coleman-Weinberg mechanism \cite{CW}.
The second term in the bracket of (\ref{MT})
is obviously  due to exchanging energy with the heat reservoir.
The low temperature limit $T \rightarrow 0$ is trivial, as $M(g,T) \rightarrow
(1+\frac{1}{6\pi} g^2)M_r > 0$.
On the other hand, since $M(g,T)$ is a monotonically
decreasing function of temperature, as $T$ goes up,
the thermal fluctuation tends to drive the effective mass
to zero. Namely, there must exist a critical temperature
$T_c$ at which  $M(g, T)=0$, and a phase transition
happens. $T_c$ is readily to solve from (\ref{MT}). We obtain  at this order
\begin{equation}
e^{-aM_r/T_c} = 1-e^{-M_r/T_c}\;,
\end{equation}
with $ a=(1/2+3\pi/g^2).$
In a linearized form, as a good approximation when $T_c \gg M_r$,
\begin{equation}
T_c \simeq 3(\frac{1}{2} + \frac{\pi}{g^2})M_r\;.
\end{equation}
Now we see that the Chern-Simons
interaction is responsible to the phase transition
as it should be, and a stronger interaction causes
a transition at a lower temperature, with the
renormalized mass $M_r$ fixed.

As is mentioned in section 2, the Chern-Simons coupling
$g$ receives only trivial correction at zero temperature
 so that the beta function of $g$ identically vanishes.
Alternatively said, it is insensitive to the change of energy scale.
However, the thermal effect at finite temperature will affect
the strength of the Chern-Simons coupling. Namely, in the
finite temperature field theory, the coupling constant is
a function of temperature, $g = g(T)$. Then it is crucial to the
perturbation expansion over the Chern-Simons coupling that
$g(T)$ must be reasonably small for $T$'s in the range $[0, T_c]$.
The effective (finite temperature) coupling constant $g(T)$
can be defined as
\begin{equation}
\Gamma_{\mu\nu\lambda}(p=0)=g(T)\epsilon_{\mu\nu\lambda}\;,
\end{equation}
where $\Gamma_{\mu\nu\lambda}(p)$ is the three-point function.
Calculating the one loop diagram for the three vertex,
 we have
\begin{equation}
\Gamma^{(2)}_{\mu\nu\lambda}(p=0)
=\frac{2}{3}g^3\epsilon_{\mu\nu\lambda}J(M,T)\;,
\end{equation}
and then the effective coupling is
\begin{equation}
g(T) = g_r\left(1-\frac{1}{6\pi}g_r^2
[1+2\frac{T}{M_r}{\rm ln}(1-e^{-M_r/T})]\right) + {\cal O}(g^5)\;,
\label{gt}
\end{equation}
where the renormalized zero-temperature coupling
$g_r = g(1 + \frac{g^2\Lambda}{3\pi^2M})$ in the regularization
by a naive cutoff, or $g_r=g$ in the dimensional regularization.
Again, in (\ref{gt}), we have replaced the bare parameters $g$ and $M$
with the renormalized ones $g_r$ and
$M_r$, and this affects only higher orders.
It shows that $g(T)$ is a monotonically slowly
increasing function of the temperature $T$,
with $g(T=0)
= g_r(1-\frac{1}{6\pi}g_r^2)$ at zero temperature. On the other hand,
at the critical temperature $T=T_c$, the effective coupling is
\begin{equation}
g(T_c) = 2g_r\;.
\end{equation}
This implies that in the region $T < T_c$, the
perturbation expansion over the
Chern-Simons coupling is  reliable,
only if the renormalized zero temperature coupling
constant $g_r$
is small.

Though the gauge fields
at zero temperature have no dynamical
mass due to the gauge symmetry,
the quantum thermal fluctuations may endow
them thermal masses. The electric and
magnetic masses
can be defined via the polarization tensor
$\Pi_{\mu\nu}(p, g,M_r,T)$ \cite{note1}:
\begin{equation}
{\cal M}_{el}(g_r,M_r,T)\delta_{\mu 0}
\delta_{\nu 0}
+{\cal M}_{mag}(g_r,M_r,T)
\delta_{\mu j}\delta_{\nu j}
= -\Pi_{\mu\nu}(p=0,g_r,M_r,T)\;.
\label{pi}
\end{equation}
Calculating the one-loop diagram for the
polarization tensor at $p=0$, we obtain
\begin{equation}
 \Pi_{\mu\nu}^{(2)}(p=0, g,M,T)
= -\frac{2}{3}g^2\delta_{\mu\nu}
\left(J(M,T)+\frac{M}{4\pi}\right)\;,
\end{equation}
and the thermal masses ${\cal M}_{el}(g_r,M_r,T)
={\cal M}_{mag}(g_r,M_r,T) = {\cal M}(g_r,M_r,T)$ with
\begin{equation}
 {\cal M}(g_r,M_r,T)
=
-\frac{2}{3\pi}g_r^2T{\rm ln}(1-e^{-M_r/T}) + {\cal O}(g^4)\;.
\label{mcs}
\end{equation}
Above we have set the renormalized zero temperature
masses of the gauge field to zero by using
counter terms when it is necessary (for instance in the
regularization by a large momentum cutoff), so that the gauge
symmetry is respected. And we have replaced the bare zero
temperature coupling $g$ and vector mass $M$ in (\ref{mcs})
with the renormalized ones.
The thermal masses of the gauge field are
monotonically increasing function of $T$.
At the critical temperature $T_c$
\begin{equation}
{\cal M}(T_c) = M_r(1+\frac{1}{6\pi}g_r^2) + {\cal O}(g^4)\;.
\end{equation}
The electric and magnetic masses ${\cal M}_{el}(T)$ and
 ${\cal M}_{mag}(T)$ are known as the inverse screening length
in the plasma \cite{GPY}\cite{K}.
Unlike those in the $(3+1)$
dimensional $QED$ and $QCD$ where
the magnetic screening is absent, our results
here suggest that both the static electric and magnetic fields are
screened by the plasma thermal excitations in the $(2+1)$ dimensional
massive system. And thus, at least to the one loop order, the plasma of
thermal excitations acts like a superconductor.

\vskip 0.5truein
\section{Zeroth Order Transition}
\setcounter{equation}{0}
\vspace{3 pt}

We turn to calculation of free energy density for the system.
As is known, the free energy is the single most important
function in the thermodynamics. From it all other thermodynamic
properties can be determined. In the present case, the free energy
may tell us the nature of the topological phase transition.
With the conventional Fourier transformation, we come to
the momentum space. At the lowest order, performing the Gaussian
integrals in (\ref{Z}), we have
the partion function
\begin{equation}
Z_0 =
[{\rm det}(-\epsilon_{\mu\nu\lambda} p_\lambda + M\delta_{\mu\nu})]^{-1}
[{\rm det}(-\epsilon_{\mu\nu\lambda} p_\lambda
+ \frac{1}{\rho}p_\mu p_\nu)]^{-\frac{1}{2}}
{\rm det}(p^2)\;.
\label{Z0}
\end{equation}
The determinants are readily calculated. The one for the CS term,
the second in the above, is
\begin{equation}
[{\rm det}(-\epsilon_{\mu\nu\lambda} p_\lambda
+ \frac{1}{\rho}p_\mu p_\nu)]^{-\frac{1}{2}} = \sqrt{\beta\rho}
\prod_n\prod_{{\bf p}}(\beta^2 p^2)^{-1}\;.
\end{equation}
The products are canceled by the determinant for ghost term, the last
in (\ref{Z0}), and the gauge parameter term $\sqrt{\beta\rho}$ contributes
the zero-point energy. This verifies
Chern-Simons gauge field carries no local dynamical degree of
freedom. On the other hand, the determinant for the free massive
$B_\mu$ field gives
\begin{equation}
[{\rm det}(-\epsilon_{\mu\nu\lambda}p_\lambda
+ M\delta_{\mu\nu})]^{-1} = \frac{1}{\beta M}
\prod_n\prod_{{\bf p}}[\beta^2 (p^2+M^2)]^{-1}\;.
\end{equation}
Then the leading contribution to the free energy density,
$ {\cal F}_0 = -{\rm ln}Z_0/(\beta V)$, is
\begin{equation}
 {\cal F}_0 = \frac{1}{2\beta V}\sum_n\sum_{\bf p}{\rm ln}
\left(\frac{\beta M^2}{\rho}\right)
 + 2T\int\frac{d^2p}{(2\pi)^2}\left[\beta\omega
+ ln(1-e^{-\beta\omega})\right]\;,
\label{F0}
\end{equation}
where $\omega = \sqrt{{\bf p}^2+M^2}$. The second term is just the
free energy of a gas of noninteracting, massive bosons.
Namely that the massive
vector $B_\mu$ obeys the Bose-Einstein distribution.
The factor ``$2$'' in the second term in (\ref{F0}) indicates
two degrees of freedom,
carried by the complex field $B_\mu$, in the thermal equilibrium.
 Besides, like the gauge fixing parameter $\rho$,
the vector mass $M$ has an extra contribution to the zero
point energy, as shown in the first
term of (\ref{F0}).

Dropping the zero-point energy, the free energy density ${\cal F}_0$ can
be re-written as
\begin{equation}
{\cal F}_0 =  -\frac{3}{\pi}T^3h_4(\frac{M_r}{T})
+ {\cal O}(g^2)\;,
\label{F01}
\end{equation}
where
\begin{equation}
h_n(x) = \frac{1}{\Gamma(n)}\int_0^\infty dy \frac{y^{n-1}}
{\sqrt{y^2+x^2}}\frac{1}{e^{\sqrt{y^2+x^2}}-1}\;.
\end{equation}

The perturbation corrections to the free energy may be calculated
by expanding the partition function (\ref{Z}) in the Chern-Simons
coupling $g$ and calculating the resulting Feynman diagrams. At
the second order, it is

\unitlength=1.00mm
\linethickness{0.4pt}
\thicklines
\begin{picture}(110.0,33.0)
\put(55.00,18.00){\makebox(0,0)[cc]{${\rm ln} Z_2 = -\frac{1}{2}$}}
\put(68.00,18.00){\circle*{2.00}}
\put(82.00,18.00){\circle*{2.00}}
\put(75.00,18.00){\circle{40.00}}
\multiput(68.0,18.0)(2.00,0.00){7}{\line(3,0){1.00}}
\end{picture}
\vspace{-0.6cm}

\noindent with the real (dashed) line
standing for the $B_\mu$ ($a_\mu$) propagator. Calculating
the two-loop diagram, we obtain
\begin{equation}
{\rm ln}Z_2 = - g^2M\beta V\left(
J(M,T)\right)^2\;,
\end{equation}
and the correction to the free energy density
\begin{equation}
{\cal F}_2 
=  \frac{1}{(4\pi)^2}g_r^2M_rT^2\left(\frac{M_r}{T}+
2{\rm ln}(1- e^{-M_r/T})\right)^2 + {\cal O}
(g^3)\;.
\label{F2}
\end{equation}
Being positive-definite, ${\cal F}_2$
increases monotonically
with $T$. Physically,
this implies the quantum thermal fluctuation
tends to decrease the
pressure $P(T) = -{\cal F}(T)$, contrary to the
thermal behavior of the system in the free theory limit,
as shown in (\ref{F01}). It would be interesting to
calculate the higher order corrections, for instance, the next
order $g^3$ from the ring diagrams. To our main purpose of
looking at the nature of the phase transition at large $T \sim T_c$,
the result upto $g^2$ is sufficient.

At the critical $T_c$, as $T_c \simeq 3(\frac{1}{2} + \frac{\pi}{g^2})M_r \gg
M_r$ for weak CS interactions, we consider the leading
term in both ${\cal F}_0$ and  ${\cal F}_2$.
It is readily to see
\begin{eqnarray}
{\cal F}_0(T_c) &=& -81\pi^2h_4(0)M^3_r\frac{1}{g^6_r}
+ {\cal O}(\frac{1}{g_r^4})\;,\\
{\cal F}_2(T_c) &=& \frac{9}{4}M^3_r\frac{1}{g^2_r} + {\cal O}(g_r^2),
\end{eqnarray}
where $h_4(0) = 2\zeta(3) \sim 2.404$.
This shows that near the critical
temperature $T_c$, ${\cal F}_0$ dominates.
In other words, the pressure of the system to the second order
is positive-definite as it approaches the critical
temperature from the below,
\begin{equation}
P(T \sim T_c) = - {\cal F}(T \sim T_c) > 0\;.
\end{equation}

Now let us look into the symmetry phase, where the mass of $B_\mu$ field
vanishes. Due to the topological nature,
the free energy must identically vanish.
This can be readily verified  within the covariant formalism
used in this work. The symmetry phase of the system is described by
the pure non-Abelian Chern-Simons theory  (\ref{Scs}). In a covariant
gauge fixing, as usual one introduces ghost term
\begin{equation}
 \partial_\mu \bar{c}^aD_\mu c^a\;,
\end{equation}
along with a gauge fixing term such as $(\partial_\mu A^a_\mu)^2/(2\rho)$.
Doing so, one fixes diffeomorphism too. This ghost term includes
the kinetic of ghost and gauge interaction
between the ghost and Chern-Simons field. The Gaussian integral of
the kinetic term of ghost $c^a$ with $a=1,2,3$ cancel that of the Chern-Simons
term for $A^a_\mu$, and so ${\cal F}_0 \equiv 0$.   And then,
order by order, the contributions to quantities like
the free energy from the ghost loops cancel out completely
those from the Chern-Simons field loops, at zero temperature
\cite{CSW} as well as in the thermal ensemble. In particular,
\begin{equation}
P(T) \equiv 0\;.
\end{equation}
This seems a natural consequence of lacking of dynamical
degree of freedom in a topological theory (or in a topological
phase of a system).

Since the  the pressure (or free energy) discontinues at the
critical temperature $T_c$, we conclude that the topological
phase transition is of the zeroth order.

\renewcommand{\theequation}{\thesection.\arabic{equation}}
\vskip 0.5truein
\section{Summary}
\setcounter{equation}{0}
\vspace{3 pt}

In conclusion, we have looked into
the thermal behavior of the
theory with a charged massive vector matter coupled to
Chern-Simons gauge field. To the second order in perturbation expansion,
 we have obtained a critical
temperature $T_c$,  at which the effective mass of vector
field vanishes, and the system transfers from a
symmetry broken phase to topological phase.
The phase transition seems to be of the zeroth order,
as the free energy of the system is discontinuous at $T_c$.
We have seen that the effective Chern-Simons coupling
$g(T)$ is a slowly increasing function
of temperature $T$ in the symmetry broken phase, and therefore
perturbation expansion over the Chern-Simons coupling
should be reliable in the whole range of $T$, provided the
renormalized zero temperature coupling is small.

The present investigation presents to our knowledge
the first example that
exhibits restoration of topological invariance
at some critical temperature. At the phase transition point,
the local dynamical degrees of freedom and spacetime metric
disappear. Meanwhile, the thermodynamical properties of the system
subject to sudden changes. For instance, the pressure of the
system promptly withdraws. And it is not difficult to
check that (all) other thermodynamical quantities derivable from
the free energy are discontinuous at the critical $T_c$ as well.
This is a phenomenon that has not been
observed in the Nature nor in theories, and deserves further studies.

It is interesting to notice that the pure Chern-Simons
theory with the non-Abelian gauge group $ISO(2,1)$
is equivalent to the $(2+1)$
dimensional general relativity with a vanishing cosmological
constant \cite{W3}.
This is realized by identifying the vierbein and spin connection
as the gauge field $A_\mu^a$, and mapping the Einstein action
to the Chern-Simons action (\ref{Scs}).
Then, in this $(2+1)$ dimensional theoretical laboratory, we
have seen a possible realization of the topological invariance
in quantum gravity. Namely that 
the thermal fluctuation plays a key role to drive the dynamical
system under consideration to a phase of pure gravity without a
spacetime metric and local
dynamical degree of freedom but with topological symmetry.

The author thanks L. Dolan, C. Itoi, G. Semenoff and Y.S. Wu
for discussions. This work was supported in part by
the  U.S. DOE under contract No. DE-FG05-85ER-40219.


\baselineskip=18.0truept

\begin{thebibliography}{199}
\bibitem{Pe} See, for instance,
R. Penrose, in {\it Magic without Magic},
J. Klauder (ed.), San Francisco Freeman, 1972.
\bibitem{A}
It was suggested by
M.F. Atiyah, in {\it The Mathematical Heritage of Hermann
Weyl}, Proc. Symp. Pure Math. {\bf 48}, R. Wells (ed.),
Providence RI:
American Mathematical Society 1988, 285-299.
\bibitem{W1} E. Witten,
Commun. Math. Phys. {\bf 117}, 353 (1988).
\bibitem{W2} E. Witten,
Commun. Math. Phys. {\bf 121}, 351 (1989).
\bibitem{MS} G. Moore and M. Seiberg,
in {\it Physics, Geometry and Topology}, H.C. Lee (ed.)
Plenum, New York, 1990) p.263.
\bibitem{W3} E. Witten,
Nucl. Phys. {\bf B}, 46 (1988).
\bibitem{Ts} A. Tseytlin,
J. Math. Phys. {\bf 15} L105 (1982).
\bibitem{DJT}
S. Deser, R. Jackiw, and S. Templeton,
Phys. Rev. Lett. {\bf 48}, 975 (1982).
\bibitem{Ho} G.T. Horiwitz,
Commun. Math. Phys. {\bf 125}, 417 (1989).
\bibitem{CH}
W. Chen, {\it Phase Transition in (2+1) Dimensional Quantum Gravity},
IFT-490-UNC (1994).
\bibitem{L} D.A. Kirzhnits and A.D. Line,
Phys. Lett. {\bf 42B}, 417 (1972).
\bibitem{DJ} L. Dolan and R. Jackiw,
Phys. Rev. D {\bf 9}, 3320 (1973).
\bibitem{W} S. Weinberg,
Phys. Rev. D {\bf 9}, 3357 (1973).
\bibitem{GPY}
D. Gross, R. Pisarski, and D. Yaffe,
Rev. Modn. Phys. {\bf 53}, 43 (1981).
\bibitem{CI1}
W. Chen and C. Itoi, Phys. Rev. Lett. {\bf 72}, 2527 (1994);
IFT-488-UNC/NUP-A-94-3.
\bibitem{JN}
R. Jackiw and V.P. Nair, Phys. Rev. D {\bf 43}, 1933 (1991).
\bibitem{Chen} W. Chen,
Phys. Rev. D {\bf 41}, 1172 (1990).
\bibitem{K}
J.I. Kapusta, {\it Finite-Temperature
Field Theory}, Cambridge Univ. Press, New York 1989.
\bibitem{AB}
A similar discussion has been conducted also
by R. Amorim and J. Barcelos-Neto,
preprint IF-UFRJ-05/94.
\bibitem{Di}
P.A.M. Dirac, {\it Lectures on Quantum Mechanics}, Ueshiva Univ,
New York, 1964.
\bibitem{Se}
A.P. Senjanovic, Ann. Phys. (N.Y.) {\bf 100} 227 (1976).
\bibitem{BRST}
E.S. Fradkin and G.A. Wilkovisky, Phys. Lett. {\bf B55} 224 (1975);
I.A. Batalin and G. Vilkovisky,  Phys. Lett. {\bf B69} 309 (1977).
\bibitem{Wein}
S. Weinber, Phys. Rev. D {\bf 10},
2445 (1974).
\bibitem{MK}
P. Morley and M. Kisslinger, Phys. Rep. {\bf 51C} (2) (1979).
  \bibitem{CW} S. Coleman and E. Weinberg,
Phys. Rev. D {\bf 7}, 1888 (1973).
\bibitem{note1} This is also valid for
dynamical and/or external electromagnetic fields, if they
are minimally coupled to the vector field $B_\mu$.
\bibitem{CSW}
W. Chen, G.W. Semenoff, and Y.-S. Wu,
Phys. Rev. D {\bf 44}, R1625 (1991);
{\it ibid} {\bf 46}, 5521 (1992).
\end{thebibliography}
\end{document}